\documentclass[final]{svjour2}
\usepackage{graphicx}
\usepackage{rotating}
\usepackage{amssymb}
\usepackage{mathptmx}
\usepackage[numbers]{natbib}

\usepackage[english]{babel}
\usepackage{graphicx}
\usepackage{epsfig}
\usepackage{amsmath}
\usepackage{graphicx}
\usepackage{upgreek}
\usepackage{float}
\makeatletter
\journalname{Journal of Low Temperature Physics}

\bibpunct{}{}{,}{s}{}{,}

\begin{document}

\newcommand{\hdblarrow}{H\makebox[0.9ex][l]{$\downdownarrows$}-}
\title{Effect of a thin AlOx layer on transition-edge sensor properties}

\author{K. M. Kinnunen, M. R. J. Palosaari and I. J. Maasilta}

\institute{Nanoscience Center, Department of Physics, P. O. Box 35, FI-40014 University of Jyv\"askyl\"a, Finland\\
\email{kimmo.m.kinnunen@jyu.fi}
}
\date{30.11.2011}

\maketitle

\begin{abstract}

We have studied the physics of transition-edge sensor (TES) devices with an insulating AlOx layer on top of the device to allow implementation of more complex detector geometries. By comparing devices with and without the insulating film, we have observed significant additional noise apparently caused by the insulator layer. In addition, AlOx was found to be a relatively good thermal conductor. This adds an unforeseen internal thermal feature to the system.

\end{abstract}

\section{Introduction}

To improve transition-edge sensor (TES) \cite{OverLord} performance, many designs have been tried, one example being the Corbino-geometry (CorTES) \cite{FSN}. The main benefits of the design are the possibility to control the TES resistance over a wide range of values and being able to model the superconducting transition analytically.
Traditionally the CorTES has shown a large excess noise component, which was originally explained by fluctuation superconductivity noise (FSN) \cite{FSN}. Our recent experiments \cite{JAP} involving the measurement of the complex impedance \cite{Z} of the devices suggest that there is a significant internal thermal fluctuation noise (ITFN) \cite{Henk} part, which may account for most of the excess noise. When trying to model the measured impedance and noise features, we find that a two-body thermal model is insufficient and often even a three-body system fails to fully fit the data. 
 
 As the CorTES differs from simple TES designs by the addition of an insulating AlOx layer on top of the TES film, with a narrow strip of the insulator extending from the TES to bulk Si, one may worry about the role of the AlOx layer. To investigate what effect it could have on the TES film itself, we set out to fabricate a standard square Ti/Au TES that would have an AlOx layer partially covering it. 

 We first fabricated a bare pixel and measured the R-T, I-V and noise properties. Then a 120 nm thick AlOx layer was fabricated by e-beam evaporation on the same pixel, and the same measurements were repeated. The AlOx layer was deposited at about $ 5\cdot10^{-6} $ mBar pressure at a rate of 0.1 nm/s and the source material was $\mathrm{Al_2O_3}$  with 99.99 \% purity and 2-4 mm grain size.

Unfortunately, the AlOx process shifted the critical temperature down by 10 mK so it is difficult to make firm conclusions from that data. We only comment that the AlOx layer seemed to smoothen and decrease the total $\alpha$, calculated from I-V as $\alpha_{tot} = (T/R)dR/dT$, but that could also be related to the same heating effect that caused the $T_C$ to shift. The noise spectra look qualitatively similar.   

\begin{figure}[ht!]
\begin{center}
\includegraphics[width=0.5\linewidth,keepaspectratio]{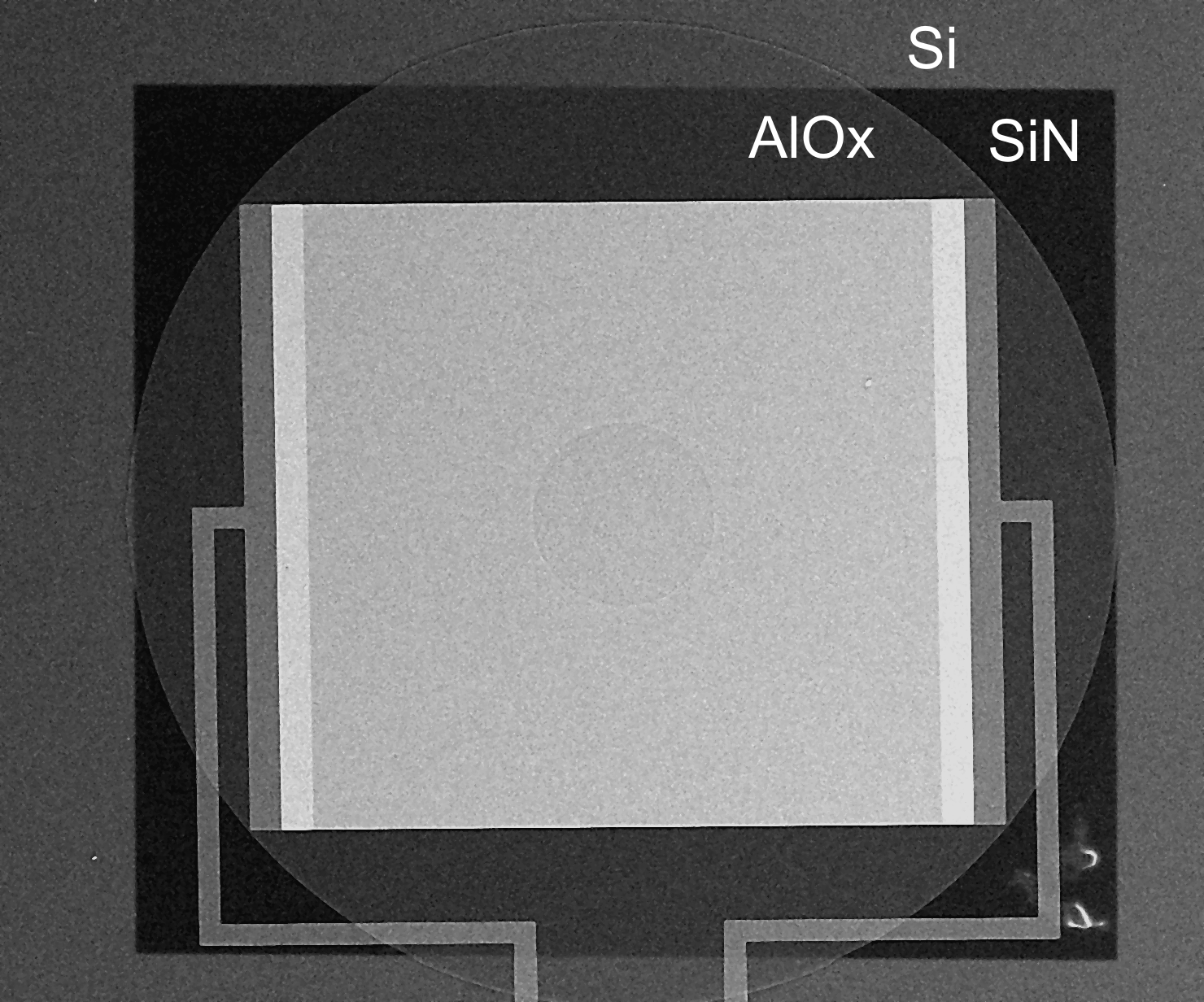}
\end{center}
\caption{ A scanning electron micrograph of the pixel A with the AlOx layer. The size of the TES is 300 x 300 $\upmu$m$^2$ and it is on a 460 x 410 $\upmu$m SiN membrane. Thickness of SiN is 750 nm.}
\label{tes}
\end{figure}

Another set of square pixels was fabricated so that this time there were identical pixels on the same chip; one was covered with AlOx while the other was left bare. We shall refer to these pixels as pixel A (with AlOx) and pixel B (bare). By accident, the AlOx layer was patterned with a wrong mask that was too large, so that it also covered most of the SiN membrane and touched on bulk Si, as shown in Fig. \ref{tes}. We chose to perform the measurements anyway and got some interesting results.

\section{Measurement results and discussion}

\begin{figure}[ht!]
\begin{center}
\includegraphics[width=0.7\linewidth,keepaspectratio]{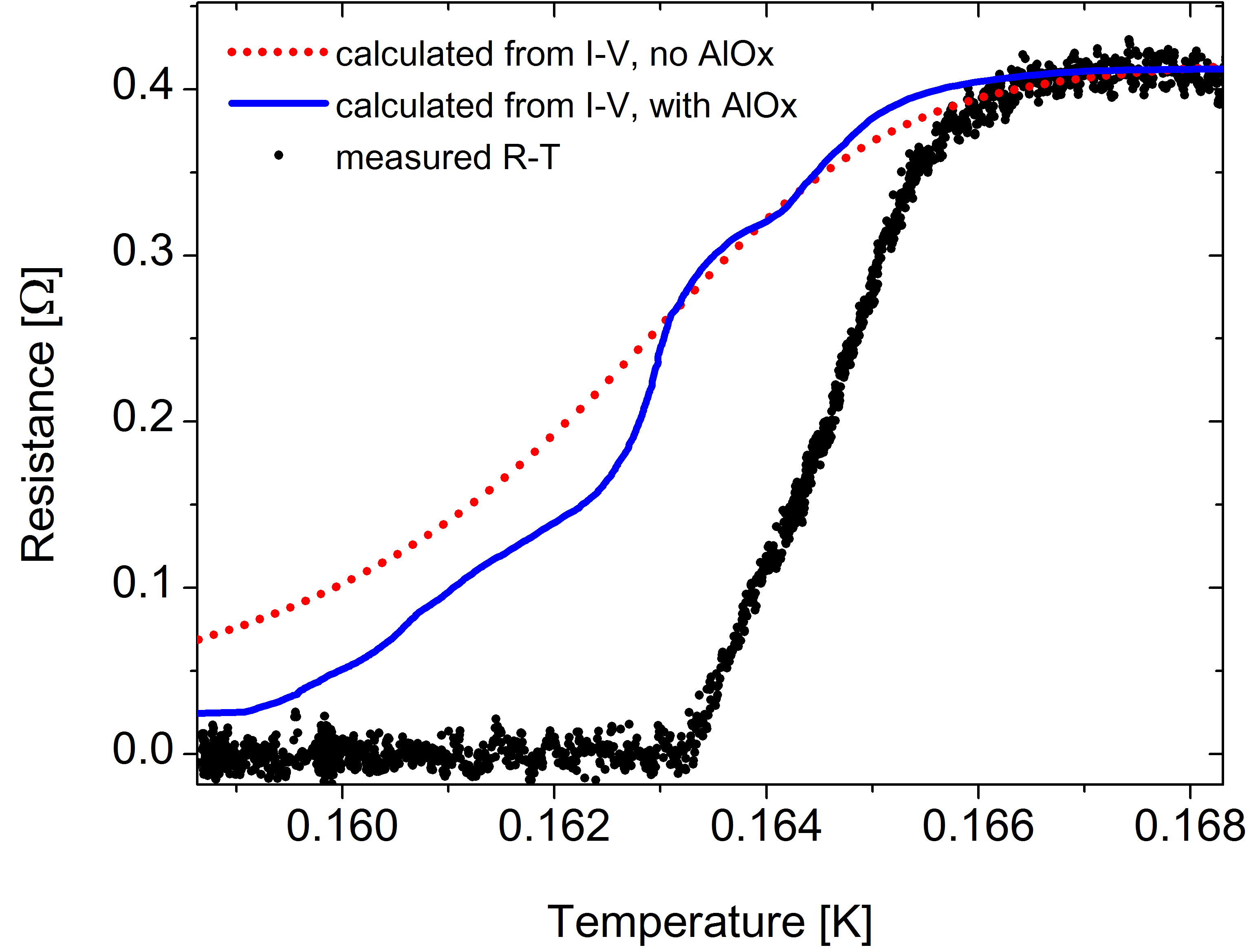}
\end{center}
\caption{(Color online) R-T curves of the pixels. Circles: four-probe lock-in measurement of pixel A. Solid and dotted lines are the R-T curves calculated from I-V measurement of pixels A and B, respectively. }
\label{RT}
\end{figure}

\begin{figure}[ht]
\begin{center}
\includegraphics[width=1.0\linewidth,keepaspectratio]{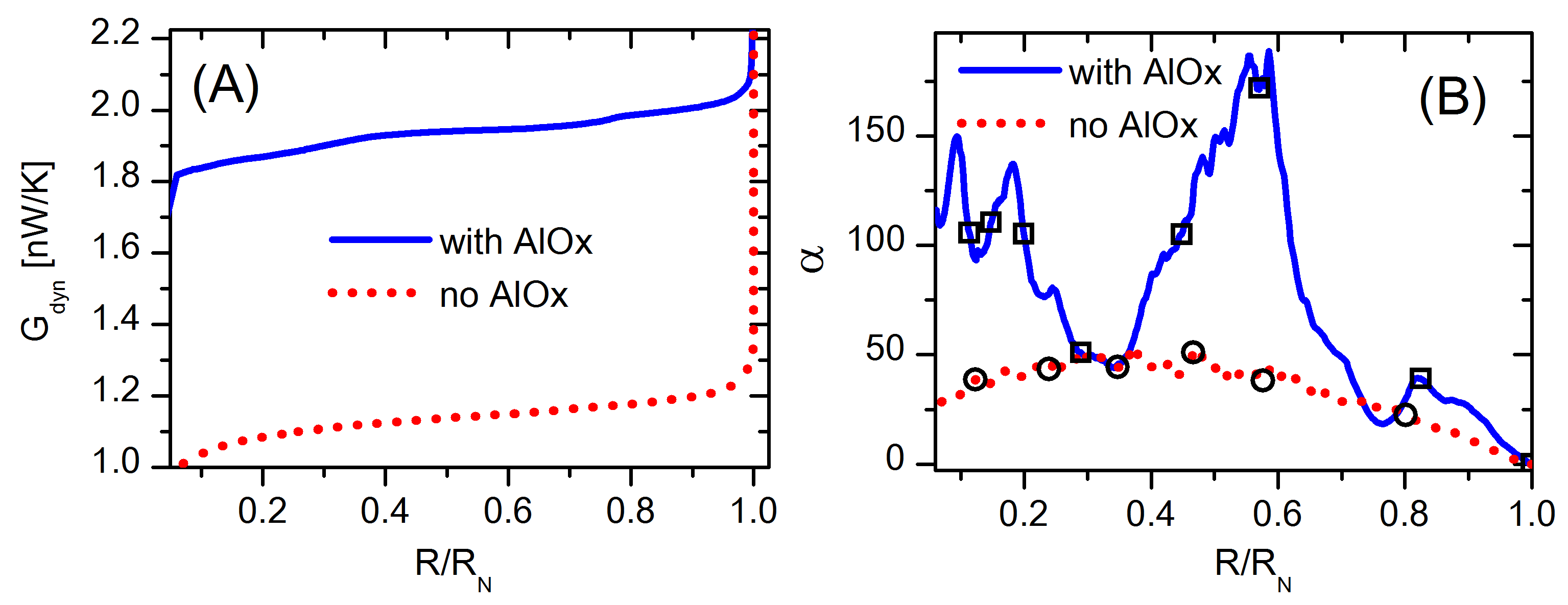}
\end{center}
\caption{(Color online) Dynamic thermal conductance (A) and total $\alpha$ (B) calculated from I-V measurement. $ G_{dyn} = dP/dT = nKT^{n-1} $, where $ P $ is the Joule heating power of the bias current, $ T $ is the TES temperature, $ n $ and $ K $ are material specific thermal transport parameters. The open symbols correspond to bias values where the noise spectra of Fig. \ref{noise} were measured.}
\label{Ga}
\end{figure}

Pixels A and B both had $T_C \approx$ 165 mK and normal state resistance $R_N \approx$ 400 m$\Omega$. As shown in Fig. \ref{RT}, pixel A has a peculiar transition shape with some very steep regions. This is reflected in the measured total $\alpha$ shown in Fig. \ref{Ga}B. Perhaps the most surprising result was the thermal conductance shown in Fig. \ref{Ga}A. While the AlOx is relatively thin compared to the SiN layer (120 nm vs. 750 nm), it almost doubles the heat transport.

\begin{figure}[ht]
\begin{center}
\includegraphics[width=0.95\linewidth,keepaspectratio]{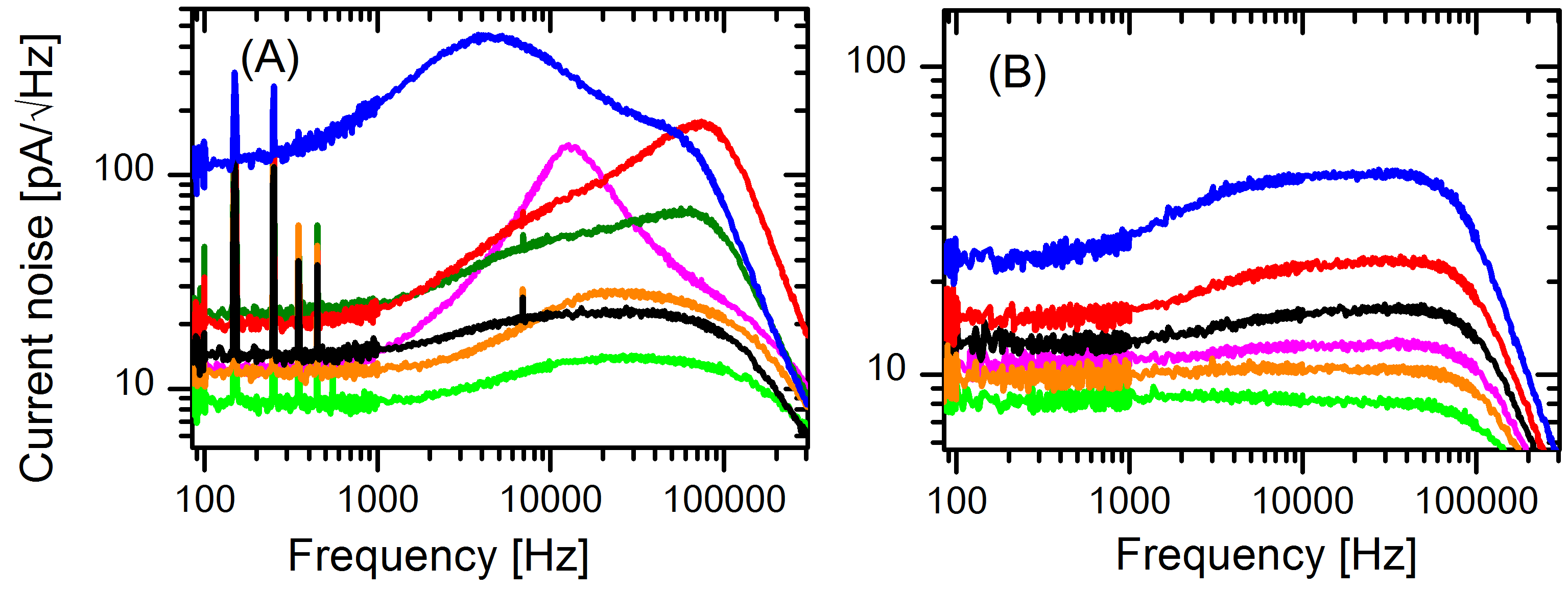}
\end{center}
\caption{(Color online) Measured electrical noise of samples with (A) and without (B) AlOx layer. The curves with same colors correspond to roughly similar bias points as shown in Fig. \ref{Ga} . Noise levels increase when going lower in the R-T curve.}
\label{noise}
\end{figure}

Figure \ref{noise} shows the measured electrical noise of the pixels. Immediately obvious are the very large noise bumps in pixel A. They are related to the high values of $\alpha$. Around the $R/R_N = 0.3$ bias point, the pixels have similar $\alpha$ but pixel A seems to have slightly higher noise level. The feature around 10 kHz is similar to what we usually see in the Corbino devices. However, in a CorTES they are not quite so pronounced and center closer to 1 kHz, possibly because the AlOx layer is a narrow strip, making thermal conductance lower. 

\begin{figure}[ht]
\begin{center}
\includegraphics[width=0.6\linewidth,keepaspectratio]{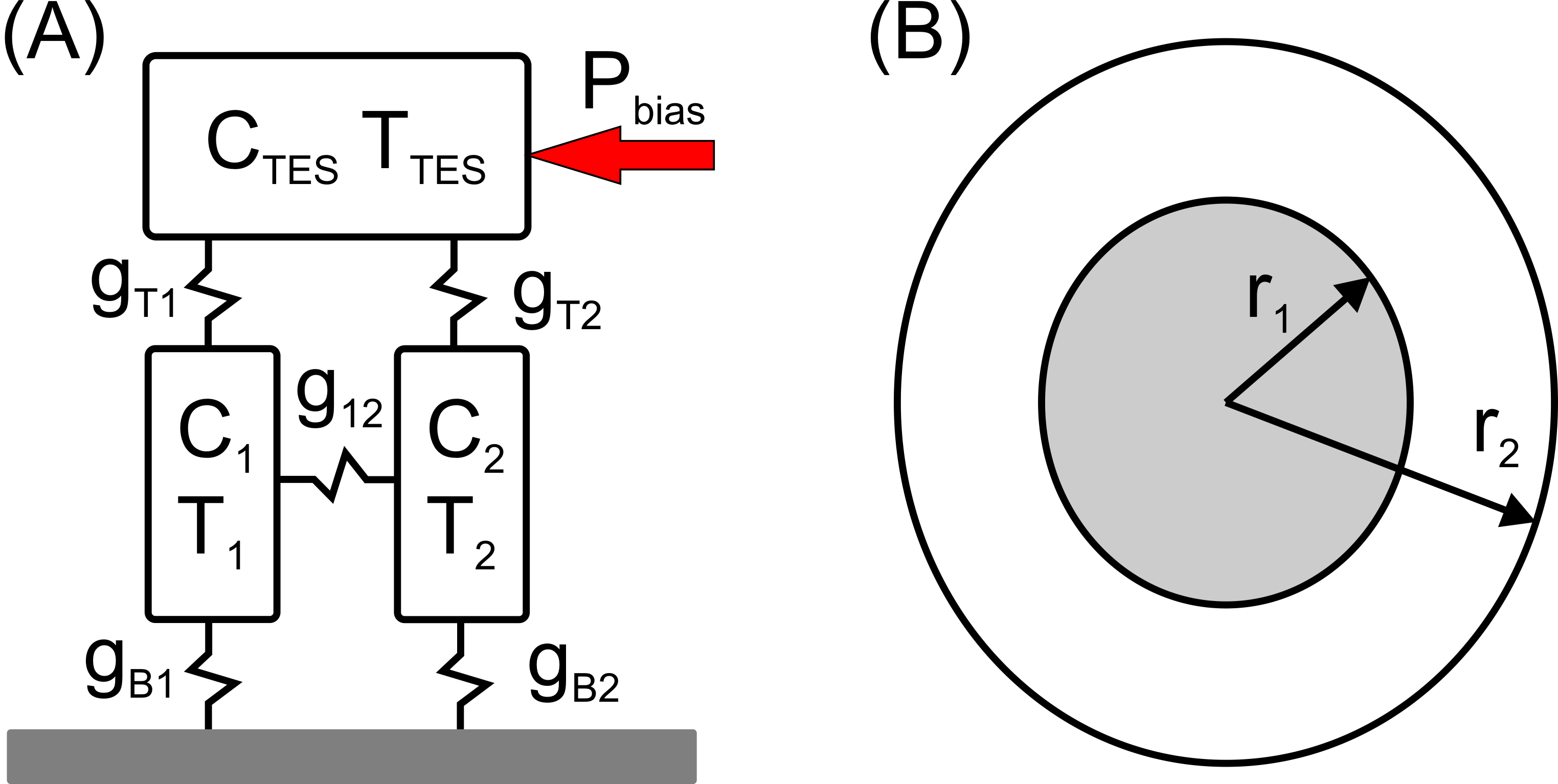}
\end{center}
\caption{(Color online) (A) Thermal block model for pixel A, not taking into account other possible thermal masses in the system. (B) Simple model used to estimate thermal conductance of AlOx. TES is approximated by a circle of radius $r_1$ = 150 $\upmu$m while the SiN and AlOx layers have radius $r_2$ = 230 $\upmu$m. }
\label{skema}
\end{figure}

We have seen in several different pixel designs that in order to explain the observed noise and complex impedance features, a thermal model where an intermediate thermal mass sits between the TES and the heat bath is usually consistent with data. It seems logical to assume that the intermediate block is related to the SiN membrane. While data from devices without AlOx layers usually fit quite well, as shown for example in ref. \cite{Mikko}, in our CorTES pixels we still have difficulties with fitting. This can be understood in light of the results presented here. The AlOx layer creates an additional conduction path to the heat bath; thus we can construct a thermal block model shown in Fig. \ref{skema}A. We therefore end up with additional features in both the noise and impedance data due to the AlOx layer. 

In the above, we have disregarded other possible additional thermal blocks, such as hanging ones. It is still unclear what is the full effect of the AlOx layer onto the device. It is possible that the AlOx film will form an additional thermal block, or it can influence the temperature profile of the TES film. The high frequency (near 100 kHz) noise bumps in pixel A hint in that direction.

\subsection{Thermal conductivity of AlOx}
We can also make a crude estimate of the thermal conductivity $\kappa$ of the AlOx layer. We assume that $\kappa = aT^2$, which is the universal temperature dependence for amorphous insulators \cite{Pohl}, where $a$ is a sample dependent constant. To derive thermal conductance $G$ from $\kappa$ we use the approximate geometry shown in Fig. \ref{skema}B to obtain \cite{Leivo} the equation

\begin{equation}
G = \frac{2\pi d}{ ln(r_2 / r_1)} \kappa,
\label{G}
\end{equation}
where $d$ is the layer thickness.

We further set $g_{12}$ in Fig. \ref{skema}A to zero so that SiN and AlOx films can be treated as parallel conductances. We can get an estimate for the thermal conductance of AlOx from I-V data by subtracting $G_{dyn}$ of pixel B from that of pixel A. We choose a value from the middle of the transition and get roughly 0.8 nW/K for AlOx. We can now calculate the thermal conductivity using eq. (\ref{G}) and get $\kappa \sim  4.5 \cdot 10^{-6}$ W/cmK for AlOx and $\kappa \sim  1 \cdot 10^{-6}$ W/cmK for SiN. The result for SiN is consistent with previous measurements \cite{Leivo}. We are not aware of 0.1 K thermal conductivity measurements for AlOx.  
According to ref. \cite{Pohl}, most amorphous glasses should have $\kappa$ in the range of $1\cdot 10^{-5} - 1\cdot 10^{-6}$ W/cmK at 0.1 K. We can therefore conclude that even though the value for AlOx we obtained is only approximate, it falls within this range and thus the layer seems to be amorphous, as expected.

\section{Conclusions}

We have shown that a thin layer of e-beam evaporated AlOx appears to be amorphous and has about five times higher heat conductivity at low temperatures than silicon nitride. We also showed that when the AlOx layer connects a TES to the heat bath, extra thermal fluctuation noise appears. This fact also lends weight to the assumption that the SiN membrane present in most devices should be taken into account when modeling the thermal circuit.

\begin{acknowledgements}
This work was supported by the Finnish Funding Agency for Technology and Innovation TEKES and EU through the regional funds, and the Finnish Academy project no. 128532. M. P. would like to thank the National Graduate School in Materials Physics for funding.
\end{acknowledgements}

\end{document}